\documentclass[journal]{IEEEtran}

\usepackage{ifpdf}
\ifpdf
  \usepackage[pdftex]{graphicx}
  \usepackage{thumbpdf}
  \usepackage[naturalnames]{hyperref}
\else
	\usepackage{graphicx}
\fi
\usepackage{bm, booktabs}
\usepackage{amsmath}%
\usepackage{amsfonts}%
\usepackage{amssymb}%
\usepackage{mathrsfs}
\usepackage{cite}
\usepackage[normalem]{ulem}
\usepackage[vlined, linesnumbered]{algorithm2e}
\newtheorem{corollary}{Corollary}

\newtheorem{lemma}{Lemma}

\newcommand{\round}[1]{\lfloor #1 \rceil}

\newcommand{\reals}{{\mathbb R}}

\newcommand{\integers}{{\mathbb Z}}

\newcommand{\Qbf}{{\mathbf Q}}
\newcommand{\Bbf}{{\mathbf B}}
\newcommand{\Ibf}{{\mathbf I}}
\newcommand{\ybf}{{\mathbf y}}
\newcommand{\xbf}{{\mathbf x}}
\newcommand{\zbf}{{\mathbf z}}
\newcommand{\ebf}{{\mathbf e}}

\newcommand{\wbf}{{\mathbf w}}
\newcommand{\kbf}{{\mathbf k}}
\newcommand{\sbf}{{\mathbf s}}
\newcommand{\Pibf}{{\mathbf \Pi}}
\newcommand{\onebf}{{\mathbf 1}}
\newcommand{\fbf}{{\mathbf f}}
\newcommand{\ubf}{{\mathbf u}}

\begin{document}

\title{An Algorithm to Compute the Nearest Point in the Lattice $A_{n}^*$}

\author{Robby~G.~McKilliam,%
  \thanks{Robby~McKilliam is partly supported by a scholarship from
    the Wireless Technologies Laboratory, CSIRO ICT Centre, Sydney,
    Australia } I.~Vaughan~L.~Clarkson%
  \thanks{Robby~McKilliam and Vaughan~Clarkson are with the School of
    Information Technology \& Electrical Engineering, The University
    of Queensland, Qld., 4072, Australia} and Barry~G.~Quinn
  \thanks{Barry~Quinn is with the Department of Statistics, Macquarie
    University, Sydney, NSW, 2109, Australia}}
% The paper headers
\markboth{Robby~G.~McKilliam \emph{et al.}, Nearest Point in the
  Lattice $A_{n}^*$}%
{\today}

% make the title area
\maketitle

\begin{abstract}
  The lattice $A_n^*$ is an important lattice because of its covering
  properties in low dimensions.  Clarkson~\cite{Clarkson1999:Anstar}
  described an algorithm to compute the nearest lattice point in
  $A_n^*$ that requires $O(n\log{n})$ arithmetic operations.  In this
  paper, we describe a new algorithm.  While the complexity is still
  $O(n\log{n})$, it is significantly simpler to describe and verify.
  In practice, we find that the new algorithm also runs faster.
\end{abstract}


\section{Introduction}

\IEEEPARstart{T}{he} study of point lattices is of great importance in
several areas of number theory, particularly the studies of quadratic
forms, the geometry of numbers and simultaneous Diophantine
approximation, and also to the practical engineering problems of
quantization and channel coding. They are also important in studying
the sphere packing problem and the kissing number problem
\cite{Clarkson1999:Anstar, SPLAG}.

A \emph{lattice}, $L$, is a set of points in $\reals^n$ such that
\[ 
  L = \{\xbf \in \reals^n |  \xbf = \mathbf{Bw} , \wbf \in \mathbb Z^n \}
\]
where $\Bbf$ is termed the \emph{generator matrix}.

The lattice $A_n^*$ is an interesting lattice due to its covering
properties in low dimensions.  It gives the thinnest covering in all
dimensions up to $8$~\cite{SPLAG}.  $A_n^*$ has also found application
in a number of estimation problems including period estimation from
sparse timing data \cite{Clarkson2007}, frequency estimation
\cite{Clarkson1999} and direction of arrival estimation
\cite{Quinn2007}.

The nearest lattice point problem is: Given $\ybf\in\reals^n$ and some lattice $L$ whose lattice points lie in~$\reals^n$, find the lattice point $\xbf \in L$ such that the Euclidean distance between $\ybf$ and $\xbf$ is minimized.  If the lattice is used for vector quantization then the nearest lattice point corresponds to the minimum distortion point.  If the lattice is used as a code for a Gaussian channel, then the nearest lattice point corresponds to maximum likelihood decoding \cite{Conway1982}. 

Conway and Sloane \cite{Conway1982} appear to have been the first to study the problem of computing the nearest lattice point in $A_n^*$.  By decomposing $A_n^*$ into a union of translations of its dual lattice $A_n$, they discovered an algorithm for computing the nearest lattice point to a given point in $O(n^2\log{n})$ arithmetic operations. Later \cite{Conway1986}, they were able to improve the execution time of the algorithm to $O(n^2)$ operations.  

Clarkson \cite{Clarkson1999:Anstar} further improved upon the work of
Conway and Sloane and described an algorithm to compute the nearest
lattice point that requires only $O(n\log{n})$ arithmetic operations.
In this paper we describe an algorithm that is similar to Clarkson's
algorithm.  Like Clarkson's algorithm, our algorithm requires
$O(n\log{n})$ arithmetic operations.  However, our algorithm and its
derivation are simpler.  The new algorithm, although of the same order
of complexity, is computationally superior.

We now describe how the paper is organized.  Section~\ref{notation}
introduces some preliminary results and definitions.  In
Section~\ref{nearestpoint} we derive all results necessary to prove
that the algorithm does find the nearest lattice point.
Section~\ref{algorithm} describes the algorithm.  A pseudocode
implementation is provided.  Is Section~\ref{complexity}, the
arithmetic complexity of the algorithm is shown to be $O(n\log{n})$.
We also tabulate some practical computation times that show the new
algorithm to be computationally superior to Clarkson's original
algorithm.

\section{Preliminary Theory} \label{notation}

Vectors and matrices are written in bold.  The $i$th element in a vector is denoted by a subscript: $x_i$.  The transpose of a vector is indicated by superscript $T$: $\xbf^T$.  We let $\bm{1}$ be a column vector of 1's and $\bm{e_i}$ be a column vector of zeros with a 1 in the $i$th position.

The \emph{Voronoi region} or \emph{nearest-neighbor region} $V(\xbf)$ of a lattice point $\xbf$ is the subset of $\reals^n$ such that, with respect to a given norm, all points in $V(\xbf)$ are nearer to $\xbf$ than to any other point in the lattice.  The Voronoi regions are $n$ dimensional polytopes~\cite{SPLAG}.  

The cubic lattice $\mathbb Z^n$ is the set of $n$ dimensional vectors with integer elements.  The Voronoi regions of $\mathbb Z^n$ are hypercubes of side length $1$.

The lattice $A_n^*$ can be defined as the projection of the cubic
lattice $\mathbb Z^{n+1}$ onto the hyperplane orthogonal to $\bm{1}$.  This is,
\begin{equation} \label{Anstar_Zn}
  A_n^* = \left\{ \Qbf \xbf \mid
        \xbf \in \mathbb Z^{n+1}  \right\} 
\end{equation}
where $\Qbf$ is the projection matrix
\begin{equation}\label{Q}
\Qbf =  \left(\Ibf - \frac{\bm{1} \bm{1}^T}{n+1}\right)
\end{equation}
where $\Ibf$ is the $(n+1)\times(n+1)$ identity matrix.

Let $\Pibf$ be a permutation matrix.  Observe the following elementary
properties:
\begin{enumerate}
\item $\Pibf \onebf = \onebf$,
\item $\onebf^T \Pibf = \onebf^T$,
\item $\|\Pibf \xbf\| = \| \xbf \|$.
\end{enumerate}

\begin{lemma} \label{communtion}
  The matrices $\Pibf$ and $\Qbf$ commute, \emph{i.e.}, $\Pibf \Qbf =
  \Qbf \Pibf$.
\end{lemma}

\begin{IEEEproof}
Using the properties of the permutation matrix, observe that
\begin{align*}
  \Pibf \Qbf &= \Pibf \big(\Ibf - \frac{\onebf \onebf^T}{n+1} \big)
        = \Pibf - \frac{\Pibf \onebf \onebf^T}{n+1} \\
  	&= \Pibf - \frac{\onebf \onebf^T}{n+1}
        = \Pibf - \frac{\onebf \onebf^T \Pibf}{n+1} \\
	&= \big(\Ibf - \frac{\onebf \onebf^T}{n+1} \big) \Pibf = \Qbf \Pibf.
\end{align*}
\end{IEEEproof}

\begin{corollary}
  For all $\zbf \in \reals^{n+1}$, $\| \Qbf \zbf \| = \| \Qbf \Pibf
  \zbf\|$.
\end{corollary}

\begin{corollary} \label{cor:x<=>Pix}
  $\xbf \in A_n^*$ if and only if $\Pibf \xbf \in A_n^*$.
\end{corollary}

\begin{IEEEproof}
  Because the inverse of a permutation matrix is also a permutation
  matrix, we need only prove sufficiency.  If $\xbf \in A_n^*$ then
  $\xbf = \Qbf \kbf$ with $\kbf \in \integers^{n+1}$.  Therefore,
  $\Pibf \xbf = \Pibf \Qbf \kbf = \Qbf \Pibf \kbf = \Qbf \kbf'$ where
  $\kbf' = \Pibf \kbf \in \integers^{n+1}$ and so $\Pibf \xbf \in
  A_n^*$.
\end{IEEEproof}

\begin{corollary}
  The lattice point $\xbf$ is a closest point in $A_n^*$ to $\ybf$ if
  and only if $\Pibf \xbf$ is a closest point in $A_n^*$ to $\Pibf
  \ybf$.
\end{corollary}

\begin{IEEEproof}
  As for Corollary~\ref{cor:x<=>Pix}, we need only show sufficiency.
  We do this by contradiction.  Suppose $\Pibf \xbf$ is not closest to
  $\Pibf \ybf$ but there is instead some $\Pibf \zbf \in A_n^*$ such
  that
  \begin{equation*}
    \| \Pibf (\zbf - \ybf) \| < \| \Pibf (\xbf - \ybf) \|
  \end{equation*}
  This implies that
  \begin{equation*}
    \| \zbf - \ybf \| < \| \xbf - \ybf \|
  \end{equation*}
  which contradicts the assumption that $\xbf$ is a closest point to $\ybf$
  in $A_n^*$.
\end{IEEEproof}

Hence, in considering an algorithm to find a closest point in $A_n^*$
to $\ybf$, it is sufficient to consider a canonical permutation of
$\ybf$.  We will see that it is very convenient to consider the
permutation in which the \emph{(centered) fractional parts of $\ybf$},
\emph{i.e.}, $\{y_i\} = y_i - \lfloor y_i \rceil$, are sorted in
descending order.  That is, in the sequel, except where otherwise
noted, we will assume that
\begin{equation}
  \{y_1\} \geq \{y_2\} \geq \dots \geq \{y_{n+1}\}.     \label{sortedy}
\end{equation}
In the case that two or more $\{y_i\}$ are equal then multiple orderings of $\ybf$ satisfy \eqref{sortedy}.  The following arguments and the subsequent algorithm are valid for any ordering of $\ybf$ that satisfies \eqref{sortedy}.

\section{Closest Point in $A_n^*$} \label{nearestpoint}

\begin{lemma} \label{lem:closest}
  If $\xbf = \Qbf \kbf$ is a closest point in $A_n^*$ to $\ybf \in
  \reals^{n+1}$ then there exists some $\lambda \in \reals$ for which
  $\kbf$ is a closest point in $\integers^{n+1}$ to $\ybf + \lambda
  \onebf$.
\end{lemma}

\begin{IEEEproof}
  Decompose $\ybf$ into orthogonal components $\Qbf \ybf$ and $t
  \onebf$ for some $t \in \reals$.  Then
\begin{equation}
  \| \ybf - \xbf \|^2 = \| \Qbf (\ybf - \kbf) \|^2 +  t^2 (n+1). \label{eq_Qdiff}
\end{equation}
  Observe that
\begin{equation*}
  \Qbf (\ybf - \kbf) = \ybf + \lambda \onebf - \kbf
\end{equation*}
where we set
\begin{equation*}
  \lambda = \frac{\onebf^T (\kbf - \ybf)}{n+1}.
\end{equation*}
Suppose $\kbf$ is not a closest point in $\integers^{n+1}$ to $\ybf +
\lambda \onebf$.  Suppose $\kbf'$ is closer.  Let $\xbf' = \Qbf
\kbf'$.  Then
\begin{align*}
  \| \ybf - \xbf' \|^2 &= \| \Qbf (\ybf - \kbf') \|^2 + t^2 (n+1) \\
        &\leq \| \ybf + \lambda \onebf - \kbf' \|^2 + t^2 (n+1) \\
        &< \| \ybf + \lambda \onebf - \kbf \|^2 + t^2 (n+1)
        = \| \ybf - \xbf \|^2,
\end{align*}
contradicting the assumption that $\xbf$ is a closest point in $A_n^*$
to $\ybf$.
\end{IEEEproof}

Now consider the function $\fbf : \reals \mapsto \integers^{n+1}$
defined so that
\begin{equation*}
  \fbf(\lambda) = \lfloor \ybf + \lambda \onebf \rceil
\end{equation*}
where $\lfloor \cdot \rceil$ applied to a vector denotes the vector in
which each element is rounded to a nearest integer\footnote{The direction of rounding for half-integers is not important.  However, the authors have chosen to round up half-integers in their own implementation.}.  That is, $\fbf(\lambda)$ gives a
nearest point in $\integers^{n+1}$ to $\ybf + \lambda \onebf$ as a
function of $\lambda$.  Observe that $\fbf(\lambda + 1) =
\fbf(\lambda) + \onebf$.  Hence,
\begin{equation}
  \Qbf \fbf(\lambda + 1) = \Qbf \fbf(\lambda).  \label{Qfl+1=Qfl}
\end{equation}

Lemma~\ref{lem:closest} implies there exists some $\lambda \in \reals$
such that $\xbf = \Qbf \fbf(\lambda)$ is a closest point to $\ybf$.
Furthermore, we see from~(\ref{Qfl+1=Qfl}) that $\lambda$ can be found
within an interval of length 1.  Hence, if we define the set
\begin{equation*}
  \mathscr{S} = \{ \fbf(\lambda) \mid \lambda \in [0, 1)\}
\end{equation*}
then $\Qbf \mathscr{S}$ contains a closest point in $A_n^*$ to $\ybf$.

If the fractional parts of $\ybf$ are sorted as in~(\ref{sortedy}), it is
clear that $\mathscr{S}$ contains at most $n+2$ vectors, \emph{i.e.},
\begin{equation} \label{scrS}
  \mathscr{S} \subseteq \big\{\lfloor \ybf \rceil,
        \lfloor \ybf \rceil + \ebf_1, \lfloor \ybf \rceil + \ebf_1 + \ebf_2,
        \dots, \lfloor \ybf \rceil + \ebf_1 + \dots + \ebf_{n+1}
 \big\}.
\end{equation}
It can be seen that the last vector listed in the set is simply
$\lfloor \ybf \rceil + \onebf$ and so, once multiplied by $\Qbf$, the
first and the last vector are identical.

An algorithm immediately suggests itself: test each of the $n+1$
distinct vectors and find the closest one to $\ybf$.  Indeed, this is
exactly the principle of the algorithm we propose here.  It only
remains to show that this can be done in $O(n \log n)$ arithmetic
operations.

\section{Algorithm} \label{algorithm}

We label the elements of $\mathscr{S}$ according to the order given
in~\eqref{scrS}.  That is, we set $\ubf_0 = \round{\ybf}$ and, for $i
= 1, \dots, n$,
\begin{equation}
  \ubf_i = \ubf_{i-1} + \ebf_i \label{eq_updateS}.
\end{equation}
Let $\zbf_i = \ybf - \ubf_i$.  Clearly, $\zbf_0 = \{\ybf\}$.
Following~\eqref{eq_Qdiff}, the squared distance between $\Qbf \ubf_i$
and $\ybf$ is
\begin{equation}
\|\ybf - \Qbf\ubf_i\|^2 = d_i + t^2(n+1) \label{fastd}
\end{equation}
where we define $d_i$ as
\begin{equation}
  d_i = \|\Qbf \zbf_i\|^2
        = \left\|\zbf_i - \frac{\zbf_i^T \bm{1}}{n + 1} \bm{1} \right\|^2
        = \zbf_{i}^T\zbf_{i} - \frac{(\zbf_{i}^T \bm{1})^2}{n+1}.
\end{equation}
We know that the nearest point to $\ybf$ is that $\Qbf \ubf_i$ which
minimizes~\eqref{fastd}.  Since the term $t^2 (n+1)$ is independent of
the index $i$, we can ignore it.  That is, it is sufficient to
minimize $d_i$, $i = 0, \dots, n$.

We now show that $d_i$ can be calculated inexpensively in a recursive
fashion.  We define two new quantities, $\alpha_i = \zbf_i^T \onebf$
and $\beta_i = \zbf_i^T \zbf_i$.  From~\eqref{eq_updateS},
\begin{equation}
  \alpha_i = \zbf_i^T \bm{1} = (\zbf_{i-1} - \ebf_i)^T \bm{1}
        = \alpha_{i-1} - 1                      \label{update_zt1}
\end{equation}
and
\begin{equation}
  \beta_i = \zbf_i^T \zbf_i = (\zbf_{i-1} - \ebf_i)^T (\zbf_{i-1} - \ebf_i)
        = \beta_{i-1} - 2\{y_i\} + 1.           \label{update_ztz}
\end{equation}

\begin{algorithm} \label{alg}
\SetAlCapFnt{\small}
\SetAlTitleFnt{}
\caption{Algorithm to find a nearest lattice point in $A_n^*$ to
  $\ybf\in\reals^{n+1}$}
\dontprintsemicolon
\KwIn{$\ybf \in \reals^{n+1}$}
$\zbf = \ybf - \round{\ybf}$ \nllabel{alg_z}\;
$\alpha = \zbf^T \onebf$ \nllabel{alg_alpha}\;
$\beta = \zbf^T\zbf$ \nllabel{alg_beta}\;
$\sbf = \textsl{dsortindices}(\zbf)$ \nllabel{alg_sortindices}\;
$D = \beta - \frac{\alpha^2}{n+1}$ \;
$m = 0$ \;
\For{$i = 1$ \emph{to} $n$ \nllabel{alg_for_all_bres}}{
 $\alpha = \alpha - 1$ \nllabel{alg_upalpha} \;
 $\beta = \beta - 2 z_{s_i} + 1$  \nllabel{alg_upbeta} \;
 \If{$\beta - \frac{\alpha^2}{n+1} < D$ \nllabel{alg_if}}{
 	$D = \beta - \frac{\alpha^2}{n+1}$ \;
 	$m = i$ \;
 }
}
$\kbf = \round{\ybf}$ \nllabel{alg_k} \;
\For{$i = 1$ \emph{to} $m$ \nllabel{alg_for_2}}{
$k_{s_i} = k_{s_i} + 1$ \;
}
$\xbf = \kbf - \frac{\bm{1}^T\kbf}{n+1}\bm{1}$ \nllabel{alg_project}\;
\Return{$\xbf$ \nllabel{alg_return}}
\end{algorithm}

Algorithm~\ref{alg} now follows.  The
main loop beginning at line~\ref{alg_for_all_bres} calculates the
$\alpha_i$ and $\beta_i$ recursively.  There is no need to retain
their previous values, so the subscripts are dropped.  The variable
$D$ maintains the minimum value of the (implicitly calculated values
of) $d_i$ so far encountered, and $m$ the corresponding index.

\section{Computational Complexity} \label{complexity}

Each line of the main loop requires $O(1)$ arithmetic computations so
the loop (and that on line~\ref{alg_for_2}) requires $O(n)$ in total. On line~\ref{alg_sortindices} the function $\textsl{dsortindices}(\zbf)$ returns the vector $\sbf$ such that $z_{s_1} \geq z_{s_2} \geq \dots \geq z_{s_{n+1}}$.  This sorting operation requires $O(n\log{n})$ arithmetic operations.  The vector operations on lines~\ref{alg_z}--\ref{alg_beta},~\ref{alg_k} and~\ref{alg_project} all require $O(n)$ operations.  It can be seen, then, that the computational cost of the algorithm is dominated by the sorting operation and is therefore $O(n \log n)$.

Clarkson's original algorithm required two sorts of $n+1$ elements.
The new algorithm requires only a single sort.  Seeing as the sort
dominates the complexity of both algorithms, we might expect our
algorithm to require approximately half the arithmetic operations of
Clarkson's original algorithm.  This appears to be the case for small
$n$.  Table~\ref{tab_computation_time} shows the practical
computational performance of Clarkson's algorithm versus our new
algorithm.  It is evident that the new algorithm is computationally
superior, particularly for small $n$.  It appears that the computational performance of the algorithms converge for large $n$.  The computer used for these trials is an Intel Core2 running at 2.13Ghz.

\begin{table}[htbp]
\centering
\caption{Computation time in seconds for $10^5$ trials}
\begin{tabular}{lrrrr}
Algorithm & \multicolumn{1}{l}{n=20} & \multicolumn{1}{l}{n=50} & \multicolumn{1}{l}{n=100} & \multicolumn{1}{l}{n=500} \\ \toprule
Clarkson & 4.57 & 6.97 & 11.11 & 47.81 \\ 
New & 2.05 & 3.86 & 7.125 & 35.44 \\ \bottomrule
\end{tabular}
\label{tab_computation_time}
\end{table}

As a final note, the algorithm proposed here can be extended to other lattices for which Lemmata \ref{communtion} and \ref{lem:closest} hold.  Potential candidates are the Coxeter lattices \cite{Coxeter1951, Martinet2003}.

\bibliographystyle{IEEEbib}

\end{document}